%% file: main.tex
\documentclass[runningheads]{llncs}
\usepackage{graphicx}
\usepackage{amssymb}
\usepackage{latexsym}
\usepackage{algorithm}
\usepackage{algpseudocode}
\usepackage{mathtools}
\usepackage{amsmath}
\usepackage{multirow}
\usepackage{booktabs}
\usepackage{wrapfig}
\usepackage{multicol}
\usepackage{comment} 
\usepackage{subcaption}
\usepackage{tabularx,booktabs}
\usepackage{wrapfig,lipsum,booktabs}
\usepackage{xcolor}
\usepackage{multirow}
\usepackage{graphicx}
\usepackage{url}
\usepackage{hyperref}

\begin{document}
\title{On the Fairness of Swarm Learning in Skin Lesion Classification}
\titlerunning{On the Fairness of Swarm Learning in Skin Lesion Classification}

\author{Di Fan\inst{1}
\and
Yifan Wu \inst{2}
\and
\thanks{Corresponding Author: Xiaoxiao Li, \href{mailto: xiaoxiao.li@ece.ubc.ca}{xiaoxiao.li@ece.ubc.ca}}Xiaoxiao Li\inst{3} 
} 
\authorrunning{Di et al.}
%
\institute{University of California, Irvine, CA, USA \and University of Pennsylvania, PA, USA \and The University of British Columbia, BC, Canada}

\maketitle             
\begin{abstract}
Fairness is essential for trustworthy Artificial Intelligence (AI) in healthcare. However, the existing AI model may be biased in its decision marking. The bias induced by data itself, such as collecting data in subgroups only, can be mitigated by including more diversified data.  Distributed and collaborative learning is an approach to involve training models in massive, heterogeneous, and distributed data sources, also known as nodes. In this work, we target on examining the fairness issue in Swarm Learning (SL), a recent edge-computing based decentralized machine learning approach, which is designed for heterogeneous illnesses detection in precision medicine. SL has achieved high performance in clinical applications, but no attempt has been made to evaluate if SL can improve fairness. To address the problem, we present an empirical study by comparing the fairness among single (node) training, SL, centralized training. Specifically, we evaluate on large public available skin lesion dataset, which contains samples from various subgroups.  The experiments demonstrate that SL does not exacerbate the fairness problem compared to centralized training and improves both performance and fairness compared to single training. However, there still exists biases in SL model and the implementation of SL is more complex than the alternative two strategies.


\end{abstract}

\input{1-intro}
\input{2-preliminaries}
\input{3-method}
\input{4-experiment}

\input{5-conclusion}

\bibliographystyle{splncs04}
\bibliography{ref}

\end{document}

%% file: 1-intro.tex
\section{Introduction}

The success of deep learning can be partially attributed to data-driven methodologies that automatically recognize patterns in large amounts of data. Machine learning can theoretically be performed locally if adequate data and computer infrastructure are available. However, data collection is costly and data sharing, such as cloud computing, is strictly regulated due to the privacy concerns in the healthcare domain. 

\begin{figure*}[!t]
    \captionsetup{justification=centering}
    \centering
    {\includegraphics[width=1\linewidth]{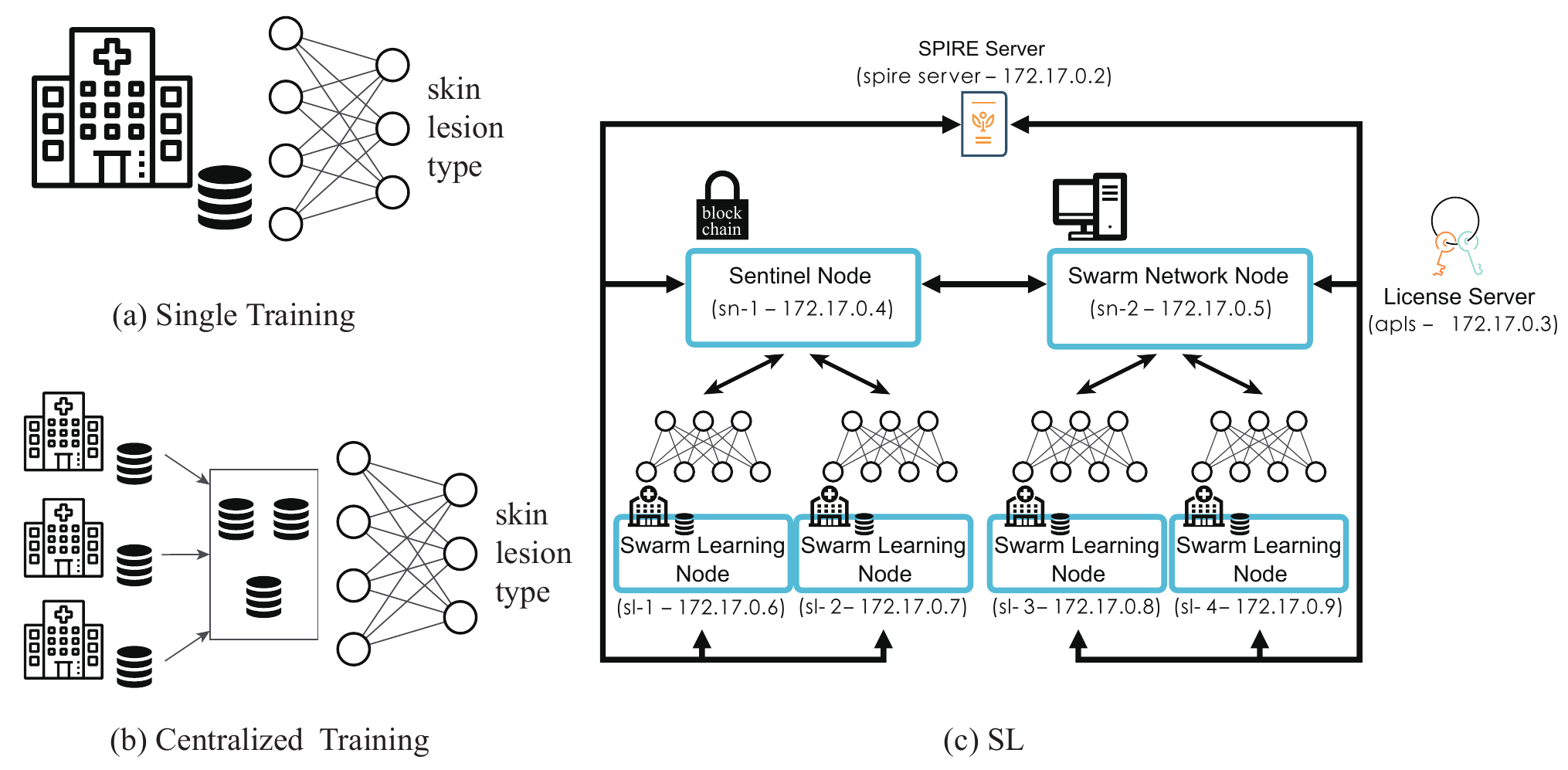}}
   
   \caption{Different training strategies to be evaluated in this study.}
    \label{fig:main}

\end{figure*}

Distributed and collaborative learning has gained popularity as a viable solution to federate disparate data and computational resources to provide a unifying analysis platform. Federated learning (FL)~\cite{mmr+17,ylct19} is a trending type of learning scheme that avoids centralizing data in model training. Local data owners (also known as nodes or clients) can train the private model locally before sending the model weights or gradients to the central server via FL. The central server then aggregates the shared model parameters to create a new global model, which it then broadcasts to each local client. Recently, a new format of collaborative learning has been proposed, called Swarm Learning (SL)~\cite{warnat2021swarm}. Unlike FL, SL replaces the central server to coordinate model updates and parameters communication with the Swarm network, which secures data privacy through blockchain technique~\cite{saito2016s} (See Figure~\ref{fig:main}(c)). Additionally, in SL, transactions can only be carried out by the nodes having been pre-authorized and new nodes can be enrolled dynamically through blockchain smart contract. Model parameters are exchanged in the Swarm network by encryption and aggregated to update a new model at each round. We will illustrate more on the SL system in Section~\ref{sec:methods}.


The statistical heterogeneity in collaborative learning and the local model multi-step training and then aggregation-based distributed learning paradigm leaves the deep learning model vulnerable to biases in the model, as the model may only attach importance to the dominant subgroup(s), such as, user groups with \textit{sensitive (protected) attributes} (\textit{i.e.} age or sex) that are over-represented in the deep learning model ~\cite{hardt2016equality,Du2020survey}. Such fairness problems have been recently discussed in the federated learning framework~\cite{li2019fair}. However, as SL is a fresh new concept proposed in recent days, no attempt has been made to evaluate the fairness or model bias issues in SL. Knowing the performance gap in different subgroups can guide researchers and practitioners to seek methods for better performance and fairness trade-off. This motivates us to conduct evaluations on the fairness in medical imaging tasks with different models. Specifically, we aim to compare model bias in centralized learning (pooling data to a data center), SL, and training model with a subgroup of data only (Figure~\ref{fig:main}). Without additional bias mitigation methods, we explore whether the fairness behavior of SL leans to centralized learning or training on the subgroup. To this end, we conduct experiments on the skin lesion classification task, as skin lesions occur in diverse populations and have a severe class imbalance. 

Our highlights are summarized in threefold:
\begin{enumerate}
    \item We implement SL for a new medical application --- skin lesion classification.
    \item To the best of our knowledge, we are the first to investigate fairness on SL.
    \item We observe that SL is robust to heterogeneous demographic-specific data distributions on our task, and it does not degrade the performance and fairness of the model compared to classical centralized training.
\end{enumerate}

%% file: 2-preliminaries.tex
\section{Related Works}
\label{sec:related}

\subsection{Collaborative learning and their application on healthcare}
Collaborative learning is a decentralized/distributed system that uses the principle of remote execution, which entails distributing duplication of a machine learning algorithm to the sites where the data is stored. These sites or devices are also called nodes. The system then runs the training method locally before sending the results to a central repository to update the algorithm~\cite{kaissis2020secure}. In the development of AI algorithms in the healthcare domain, data is one of the most important factors. However, there are two long-standing data-related challenges: 1) There are no standardized patient records; 2) Patient data is subject to stringent rules and protection standards. As a result, the concept of federated learning has recently gained a lot of attention. In the field of medical imaging, federated learning's applicability and benefits have been explored, such as on the problem of whole-brain segmentation and brain tumor segmentation~\cite{rieke2020future}, as well as finding disease-related biomarkers on fMRI~\cite{li2020multi}.

\subsection{Security and privacy of federated learning}
In federated learning, security and privacy are critical, for example, in the communication process between nodes. The existing methods include 1. Differential privacy~\cite{dwork2014algorithmic}, which is a method of preserving a dataset's global statistical distribution while eliminating individually identifiable data; 2. Homomorphic encryption, which is a type of encryption that allows unencrypted data to be processed as if it were unencrypted data; 3. Secure multi-party computation~\cite{zhao2019secure}, meaning the process is done on encrypted data shares that are distributed among them in such a way that no single party can get all of the data on their own.

\subsection{Fairness}
Data inherently is biased. Deep learning algorithms have a tendency to mimic, if not increase, data bias. For example, existing chest X-ray classifiers were discovered to have differences in performance between subgroups defined by distinct phenotypes~\cite{seyyedkalantari2020chexclusion}. Using gender-balanced datasets, a recent study found statistically significant disparities in performance on medical imaging-based diagnosis~\cite{larrazabal2020gender}. Darker skinned patients may be underrepresented~\cite{Kinyanjui20MICCAI,li2021estimating} in existing dermatology datasets~\cite{tschandl2018ham10000,codella2018skin}, similar to forms of prejudice associated to face recognition~\cite{wu2019privacy}. 
The effects of decision-making that is (partially) based on the values of biased qualities can be irrevocable or even lethal, especially in medical applications.

%% file: 3-method.tex
\section{Problem Setting and Methods}
\label{sec:methods}

\subsection{Problem Setting}
In this paper, we focus on skin lesion classification. Skin cancer is a server public healthcare problem. In the United States, over 5 million newly diagnosed cases every year~\cite{jerant2000early}. Early detection of skin lesions will decrease the death rate and reduce medical costs. Dermoscopy is one of the most widely used skin imaging techniques to distinguish the lesion spots on skin~\cite{binder1995epiluminescence}. Our classification task is based on dermoscopic images. 

Specifically, we cluster the dermoscopic images into subgroups based on patients' age and sex, which mimic the demographic-specific data distributions in different data silos. We consider data is distributed and assume each institution only contains a certain subgroup of dermoscopic images. We evaluate the model performance and fairness on unseen testing data set. To investigate model performance and bias in SL, we conduct experiments on skin lesion classification tasks in SL, and compare it to centralized training (pulling data together) and signal node training. 

Next, we will introduce the different training strategies.

\subsection{Swarm Learning}

\textbf{Swarm Learning(SL)} is a decentralized, privacy-preserving Machine Learning framework. It utilizes a dedicated server to train the Machine Learning models with distributed data sources in a blockchain-safe strategy. As shown in Figure~\ref{fig:main}(c), a set of local nodes process their local training data respectively without sharing with each other to obtain an ML model collaboratively. Then, via a Swam network rather than a centralized server, the ML parameters or weights are shared by each individual node, and finally, a merging model is formed.

At the beginning of SL, each local node should enroll or register with a blockchain Swarm smart contract. Then, this one-time process could enable each node to record information from the contract, like obtaining the model and performing local training as soon as it meets the synchronization condition. Next, the Swarm application programming interface (API) makes parameters exchanging among nodes and merges to update a new model before a new round of training. Finally, after the final round of parameter sharing and merging, the framework will check the stop criterion and halt if the criterion is reached otherwise back to local training.

With the blockchain techniques, SL has the following characteristics and advantages: (1) Saving large medical data locally; (2) No need to exchange original data, thereby reducing data traffic; (3) Providing high-level data security and protecting model from attacks; (4) No need for a secure central network; and (5) Allowing all members to merge parameters with equal rights.

\subsection{Local and Centralized Training}
Now, we describe two conventional training strategies to be compared with SL.

\textbf{Single (Node) Training} assumes no inter-institutional model and data sharing. The institution uses the local data to train deep learning models. 

\textbf{Centralized Training} collects the distributed data in the central database or cloud server and trains deep learning models following the classical approach.

\subsection{Fairness Definition and Metrics}
\label{sec:notation}
In this section, we introduce the notation of fairness in classification model and the fairness evaluation metrics to be used in this study.

 
\begin{definition}[Fair classifier] Let denote binarized sensitive attribute $z \in \{0,1\}$ that induces bias or unfairness. We define a classifier $f$ with respect to data distribution $P$ on $\{(X,Z,Y): \mathbb{R}^d \times \mathcal{Z} \times \mathcal{C}\}$, where $\mathcal{C}=\{1,2,\dots,C\}$ and $C$ is the number of classes. The classifier $f$ is fair if 
\begin{align*}
    P(f(x,z)\in \mathcal{C}|z=0) = P(f(x,z)\in \mathcal{C}|z=1)
\end{align*}
\end{definition}
Namely, if the distance between the two output distributions of $z=0$ and $z= 1$ is zero, then $f$ is claimed as a fair classifier.

Before introducing the fairness metrics, we first give the formula of true positive rate (TPR) and false positive rate (FPR). Given $x_i, y_i, z_i$ as input, label, and bias indicator for the $i$th sample in the dataset $\mathcal{D} = \{X,Y,Z\}$. We denote the data samples in class $c \in \mathcal{C}$ as $\mathcal{D}_c$. Let $\hat{y}_i$ represent the predicted label of sample $i$. 

The widely used bias quantification metrics are \textit{Statistical Parity Difference (SPD), and Equal opportunity difference (EOD)}~\cite{bellamy2018ai,dwork2012fairness,hardt2016equality}:

\paragraph{SPD} measures the difference in the probability of positive outcome between the privileged and under-privileged groups:
\begin{align*}
    SPD = \frac{1}{C}\sum_c^C \left( P_{(x_i,y_i,z_i)\in \mathcal{D}_c}(\hat{y}_i=y_i|z_i=0) - P_{(x_i,y_i,z_i)\in \mathcal{D}_c}(\hat{y}_i=y_i|z_i=1)\right ).
\end{align*}

\paragraph{EOD} measures the difference in TPR for the privileged
and under-privileged groups. TPR for a given class $c \in \mathcal{C}$ and subgroup $z$ are defined as $TPR_{z}^c = P_{(x_i,y_i,z_i)\in \mathcal{D}_c}(\hat{y}_i=y_i|z_i=z)$.

\begin{align*}
   EOD =  \frac{1}{C}\sum_c^C \left( TPR_{z=0}^c- TPR_{z=1}^c \right ).
\end{align*}




%% file: 4-experiment.tex
\section{Experiment and Results}
\label{sec:expereiment}

\subsection{Dataset}
Our data comes from the public skin lesion analysis dataset, Skin ISIC 2018~\cite{tschandl2018ham10000,codella2018skin}. However, we will divide the data into distributed portions to mimic multi-institutional data collection. The whole ISIC 2018 dataset consists of 327 actinic keratosis (AKIEC), 514 basal cell carcinoma (BCC), 115 dermatofibroma (DF), 1113 melanoma (MEL), 6705 nevus (NV), 1099 pigmented benign keratosis (BKL), 142 vascular lesions (VASC) smaples and in total 10015 RGB images. We resize the images to $224 \times 224$.  We consider sensitive attributes of ISIC dataset are \textit{age} ( $\geq$60 and $<$60), and \textit{sex} (male and female). 

\begin{figure*}[!t]
    \captionsetup[subfigure]{justification=centering}
    \centering
    \subfloat[\small Training Set ]{\includegraphics[width=0.5\linewidth]{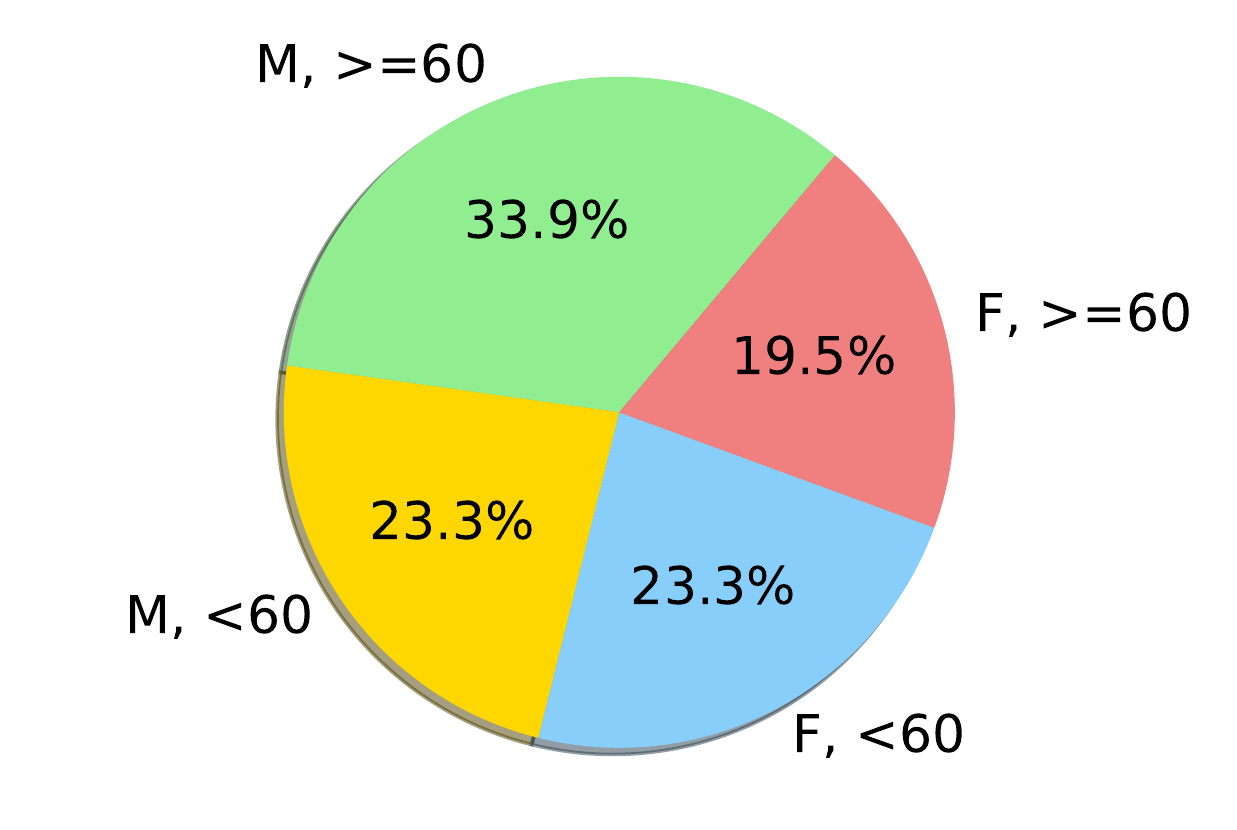}}
    \subfloat[\small Testing Set]{\includegraphics[width=0.5\linewidth]{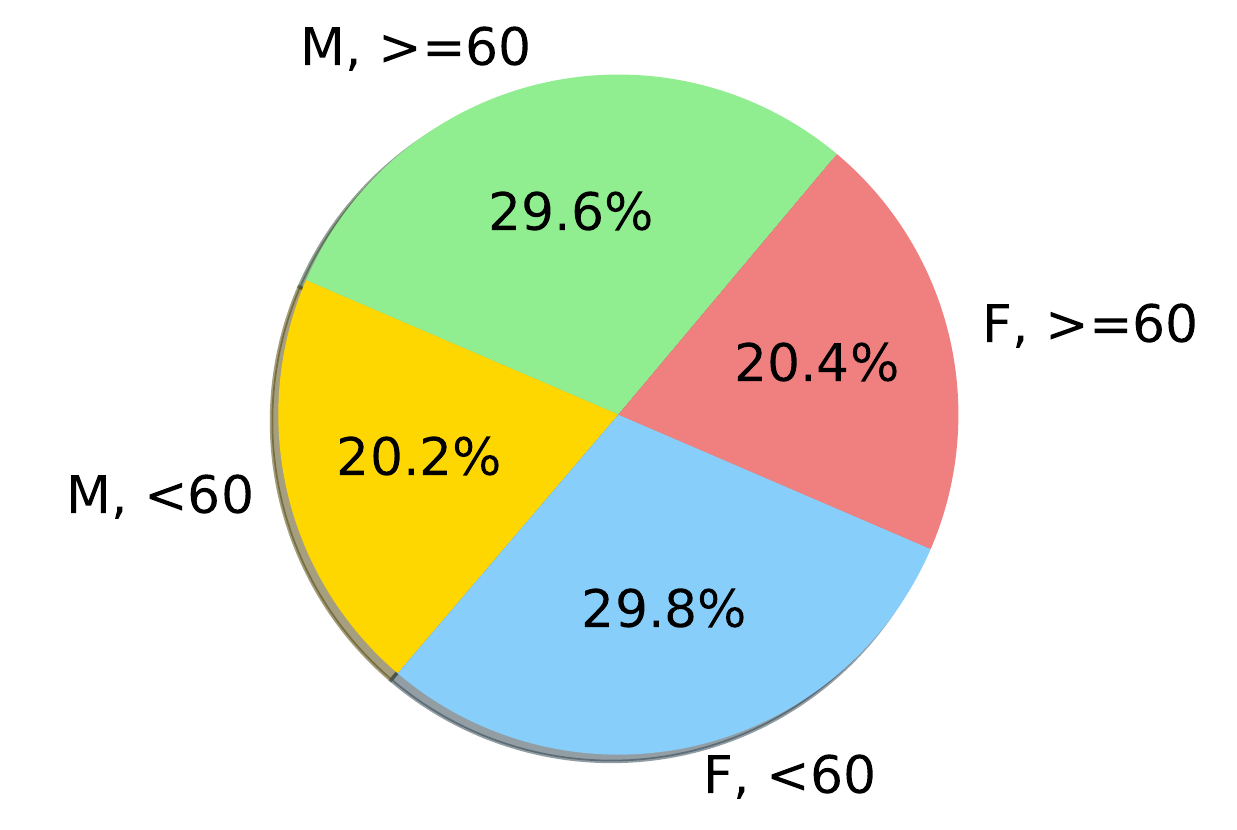}}
    \caption{\small Training data distribution (left), Test data distribution (right). The subgroup type is denoted next to the pie chart, where `M' means `male', `F' means `female', and ages are separated by $< 60$ and $\geq 60$.
    }
    \label{fig:fig1}

\end{figure*}

\begin{figure}[t!p]
         \begin{subfigure}[b]{0.5\textwidth}
                 \centering
                 \includegraphics[width=0.8\linewidth]{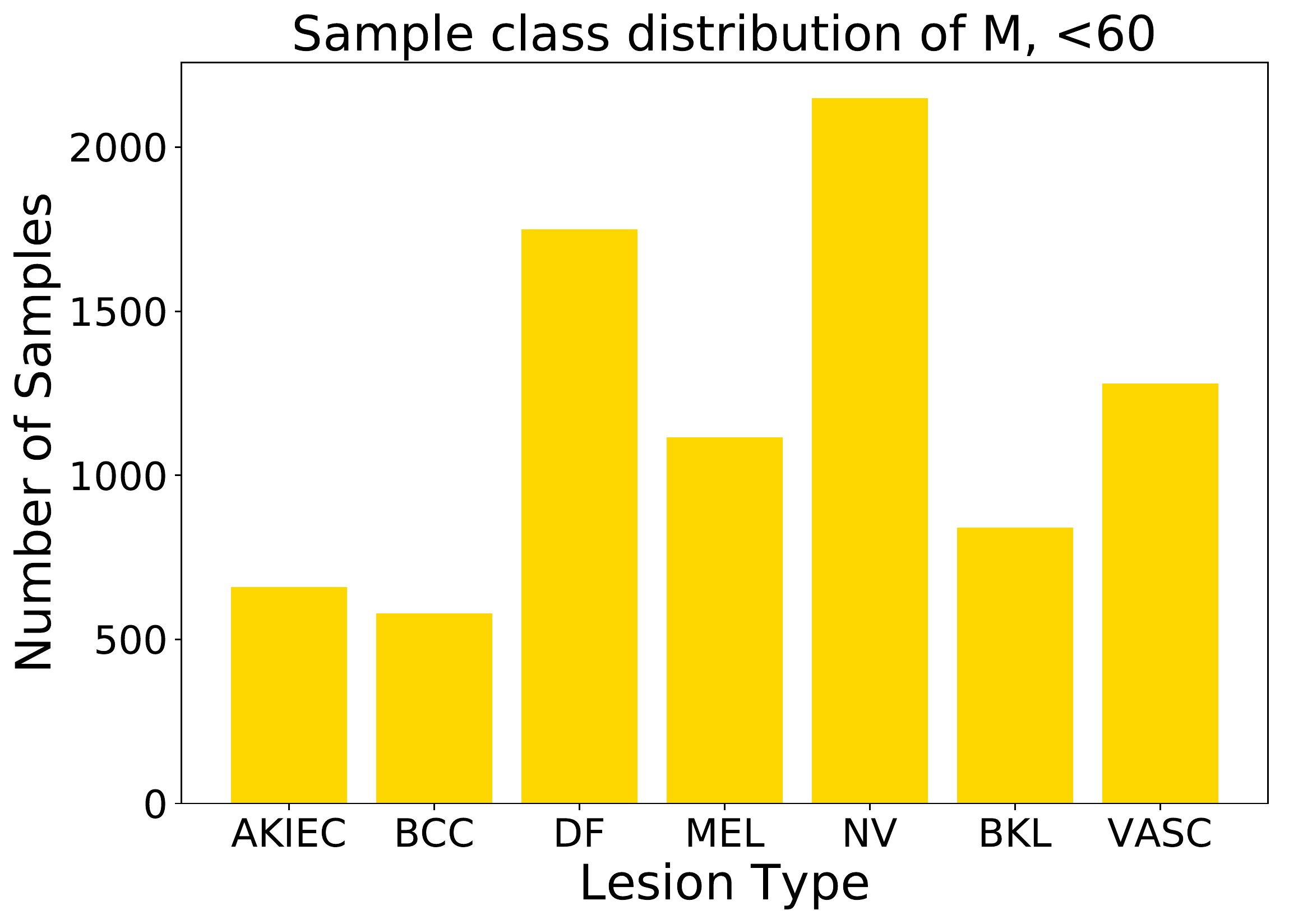}
                 \caption{Sample class distribution of M, $<$60}
                 \label{fig2:coolcat}
         \end{subfigure}
         \begin{subfigure}[b]{0.5\textwidth}
                 \centering
                 \includegraphics[width=0.8\linewidth]{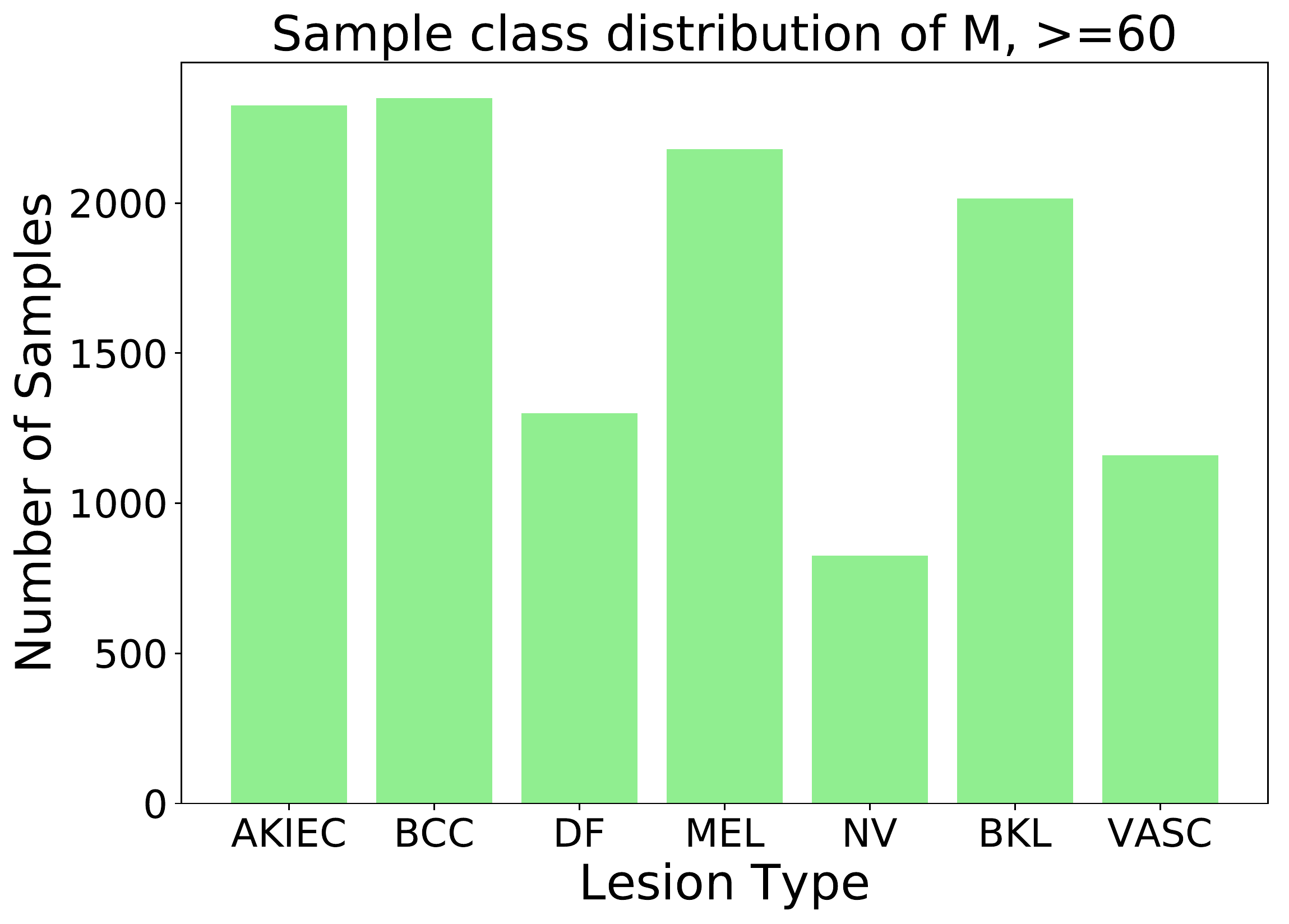}
                 \caption{Sample class distribution of M, $>=$60}
                 \label{fig2:bossycat}
         \end{subfigure}

         \begin{subfigure}[b]{0.5\textwidth}
                 \centering
                 \includegraphics[width=0.8\linewidth]{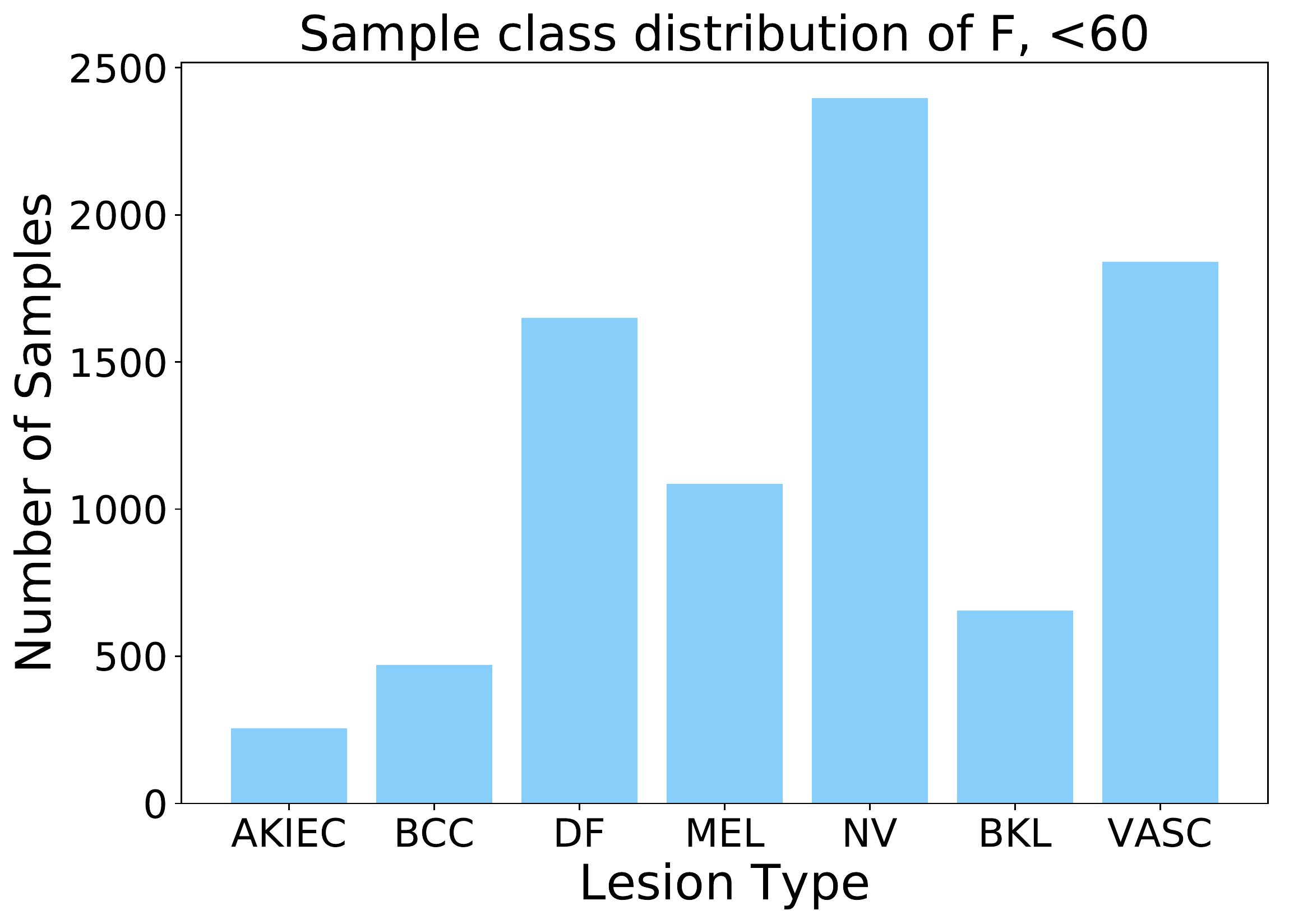}
                 \caption{Sample class distribution of F, $<$60}
                 \label{fig2:frowningcat}
         \end{subfigure}
        \begin{subfigure}[b]{0.5\textwidth}
             \centering
             \includegraphics[width=0.8\linewidth]{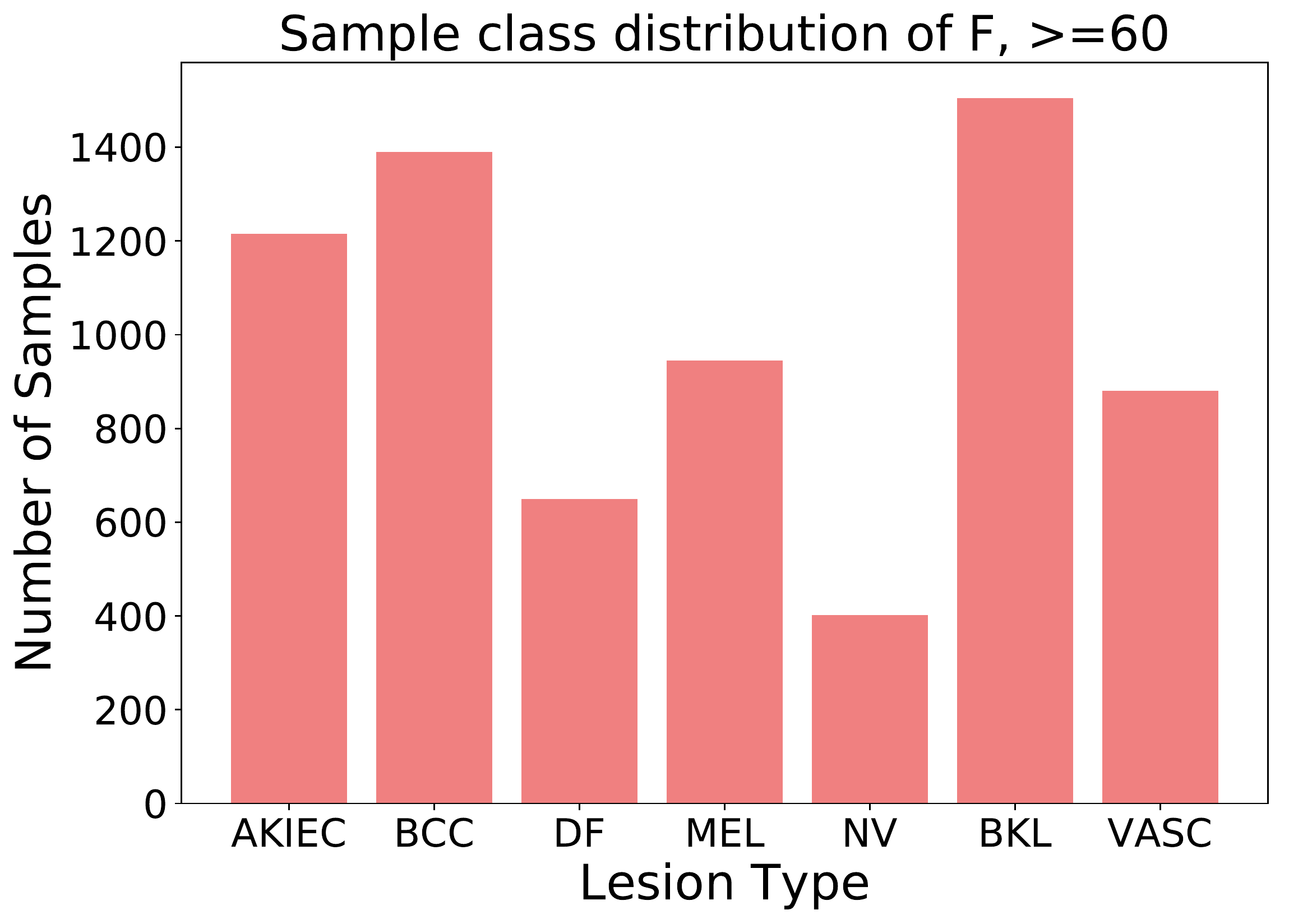}
             \caption{Sample class distribution of F, $>=$60}
             \label{fig2:scared}
     \end{subfigure}
    
\caption{Label distributions in the four subgroups (institutions).} 
\label{fig:fig2}
\end{figure}

To evaluate our algorithms, we randomly split the dataset into a training set (80\%) and a test set (20\%). As the data is unbalanced across lesion classes, we augment $15\times, 10\times, 5\times, 50\times, 0\times, 40\times, 5\times$ for AKIEC, BCC, DF, MEL, NV, BKL, and VASC, respectively. 
We divide the training data into four subgroups -- Male and age $\geq$ 60; Male and age $<$ 60; Female and age $\geq$ 60; Female and age $<$ 60. Hence we model four institutions, and each owns the data from a subgroup. After augmentation, the population proportion of the four subgroups are shown in Figure \ref{fig:fig1}(a). For testing dataset, to evaluate the model more efficiently, we make the data nearly satisfy the condition of age $<$ 60: age $>$ 60 = 1:1 and male: female = 1:1 without augmentation over the four subgroups are shown in Figure \ref{fig:fig1}(b). Finally, figure \ref{fig:fig2}  presents the heterogeneous sample distributions of the seven lesion types in the four subgroups (here, the four institutions).

\subsection{ Implementation details}
We use VGG19\cite{simonyan2014very} with cross-entropy loss to classify skin lesions among the 7 classes (AKIEC, BCC, DF, MEL, NV, BKL, and VASC). Batch size is set to 32. 
We use Adam optimizer, learning rate (lr) for classification training is 1e3 for first 10 epochs, and 1e4 after 10 epochs.

\textbf{For SL}, we use a cluster of V100 GPUs to achieve our SL framework implemented on Tensorflow 2.2. Eight individual IP addresses correspond to one License Server, one Spire Server, two Swarm Network nodes, and four Swarm Learning Nodes. Here, four Swarm learning Nodes represent four hospitals. Each Swarm learning node has four kinds of data, respectively. All Swarm Network and Swarm Learning nodes connect to the License Server, apls, running on host 172.17.0.3. All Swarm Network and Swarm Learning nodes connect to a single SPIRE Server running on host 172.17.0.2. Two Swarm Network nodes utilized are sn-1 and sn-2. sn-1 is the Sentinel Node running on host 172.17.0.4, and sn-2 runs on host 172.17.0.5. Each Swarm Network node has two Swarm Learning nodes connected to it — Swarm Learning nodes sl-1 and sl-2 connect to sn- 1 while sl-3 and sl-4 connect to sn-2. sl-1 runs on host 172.17.0.6, sl-2 runs on host 172.17.0.7, sl-3 runs on host 172.17.0.8, sl-4 runs on host 1172.17.0.9.

\textbf{For single training}, model is trained on one V100 GPU using only one subgroup of data (M, age$<60$) representing a hospital.  

\textbf{For centralized training}, we use the whole training dataset by pooling the four hospitals in SL together and train the data with one V100 GPU. 
\begin{table}[!t]
\centering
\caption{\small  Model performance. Higher scores indicate better performance.}
\label{tab:classify}
\resizebox{0.95\textwidth}{!}{%
\begin{tabular}{l|c|ccc|c|ccc}
\hline
 & \textbf{Sex} & Precision & Recall & F1-score & \textbf{Age} & Precision & Recall & F1-score    \\ \hline
\multirow{2}{*}{Centralized} & Male & 0.774 & 0.680 & 0.707 & $\geq 60$ & 0.812 & 0.752 & 0.776    \\
 & Female & 0.899 & 0.893 & 0.892 & $ < 60$ & 0.839 & 0.900 & 0.834 \\ 
 \hline
\multirow{2}{*}{\begin{tabular}[c]{@{}l@{}}SL \end{tabular}}  & Male & 0.691 & 0.658 & 0.656 & $\geq 60$ & 0.760 & 0.698 & 0.712   \\
 & Female & 0.758 & 0.731 & 0.737  & $ < 60$ & 0.638 & 0.762 & 0.660  \\ \hline
\multirow{2}{*}{\begin{tabular}[c]{@{}l@{}}Single \end{tabular}}  & Male & 0.579   &  0.460   &  0.498 & $\geq 60$ & 0.578  &   0.428  &   0.471  \\
 & Female & 0.786   &  0.547   &  0.617 & $ < 60$ & 0.761  &  0.723  &   0.716   \\ \hline
\end{tabular}%
}

\end{table}

\begin{table}[t]
\centering
\caption{\small Fairness scores. Lower SPD and EOD indicate less bias.
}
\label{tab:pred}
\resizebox{0.7\textwidth}{!}{%
\begin{tabular}{l|c>{\centering\arraybackslash}p{1.2cm}c|c>{\centering\arraybackslash}p{1.2cm}c}
\toprule
 & \multicolumn{3}{c}{\textbf{Sex} } & \multicolumn{3}{|c}{\textbf{Age}}  \\ \cline{2-7} 
 & Centralized& SL & Single  & Centralized& SL & Single\\ \hline
SPD & 0.125  & 0.067  & 0.207 & 0.027  & 0.122  & 0.183  \\ \hline
EOD & 0.213  & 0.073 & 0.087 & 0.148 & 0.064 & 0.295 \\ \bottomrule
\end{tabular}%
}

\end{table}

\subsection{Biases in Models Trained with Different Strategies}

We evaluate our framework on skin lesion detection against sensitive attribute: \textit{age} and \textit{sex}. The testing dataset contains samples from all the above four subgroups. We present the bias that existed in the centralized training, single training, and SL by showing the performance metrics (precision, recall, F1-score). The results of centralized training are listed in the \textit{`Centralized'} row of Table \ref{tab:classify}. Row \textit{`SL'} and row \textit{`Single'} list the results for SL and single training respectively. The averaged precision, recall, F1-score are reported to evaluate the testing classification performance on the subgroups with different sensitive attributes. Centralized training achieved the best results. The result is within our expectation as the distributed optimization method in SL may hurt model performance, especially on heterogeneous data. Nevertheless, SL still outperforms single training and generalizes much better to testing data that contains unseen subgroups, on the average performances of both subgroups.

We notice the performance gaps between the subgroups exist for all the investigated training strategies, which reveals the biases. We denote the group with better classification performance (such as female, $\rm age<60$) as privilege group $z=0$ and the opposite group (such as male, $\rm age\geq60$) as under-privileged group $z=1$.  We quantitatively measure the fairness scores (SPD, EOD) on the testing set for our models and the vanilla model, as shown in Table \ref{tab:pred}. Lower
fairness scores indicate the model is less biased.  We observe that SL achieves best results for fairness scores on `sex' partition and EOD for `age' partition, whereas the SPD for `age' partition is less than centralized. In this case, SL achieves comparable performance and the same level of fairness in the heterogeneous demographic-specific data setting as centralized training. Also, centralized has a higher EOD for `sex' partition than single and this could be in an explanation that single training has a poor performance for both `male' and `female' recall value in Table \ref{tab:classify} and causes the difference to be small thus makes EOD lower compared to centralized.

%% file: 5-conclusion.tex
\section{Discussion and Conclusion}
In this work, we make the first attempt to evaluate the model fairness in SL for medical applications. Specifically, we compare model performance and fairness with various metrics on single, centralized, and SL. Based on our experimental observation, SL could achieve better performance than training on a single institution, and the SL model does not amplify biases. However, we cannot conclude that SL gains the best performance-fairness trade-off for arbitrary clinical tasks. We also want to point out the high implementation complexity of SL framework due to the intricate blockchain network configurations. In summary, our work brings the interesting collaborative learning formula to skin cancer classification and designs experiments to evaluate its fairness medical applications and highlight. Our future work includes improving model performance for SL, investigating the mechanism of SL in handling model fairness, and designing bias mitigation methods for SL. 

%% file: main.bbl
\begin{thebibliography}{10}
\providecommand{\url}[1]{\texttt{#1}}
\providecommand{\urlprefix}{URL }
\providecommand{\doi}[1]{https://doi.org/#1}

\bibitem{bellamy2018ai}
Bellamy, R.K., Dey, K., Hind, M., Hoffman, S.C., Houde, S., Kannan, K., Lohia,
  P., Martino, J., Mehta, S., Mojsilovic, A., et~al.: Ai fairness 360: An
  extensible toolkit for detecting, understanding, and mitigating unwanted
  algorithmic bias. arXiv preprint arXiv:1810.01943  (2018)

\bibitem{binder1995epiluminescence}
Binder, M., Schwarz, M., Winkler, A., Steiner, A., Kaider, A., Wolff, K.,
  Pehamberger, H.: Epiluminescence microscopy: a useful tool for the diagnosis
  of pigmented skin lesions for formally trained dermatologists. Archives of
  dermatology  \textbf{131}(3),  286--291 (1995)

\bibitem{codella2018skin}
Codella, N.C., Gutman, D., Celebi, M.E., Helba, B., Marchetti, M.A., Dusza,
  S.W., Kalloo, A., Liopyris, K., Mishra, N., Kittler, H., et~al.: Skin lesion
  analysis toward melanoma detection: A challenge at the 2017 international
  symposium on biomedical imaging (isbi), hosted by the international skin
  imaging collaboration (isic). In: 2018 IEEE 15th International Symposium on
  Biomedical Imaging (ISBI 2018). pp. 168--172. IEEE (2018)

\bibitem{Du2020survey}
{Du}, M., {Yang}, F., {Zou}, N., {Hu}, X.: Fairness in deep learning: A
  computational perspective. IEEE Intelligent Systems pp.~1--1 (2020)

\bibitem{dwork2012fairness}
Dwork, C., Hardt, M., Pitassi, T., Reingold, O., Zemel, R.: Fairness through
  awareness. In: Proceedings of the 3rd innovations in theoretical computer
  science conference. pp. 214--226 (2012)

\bibitem{dwork2014algorithmic}
Dwork, C., Roth, A., et~al.: The algorithmic foundations of differential
  privacy. Found. Trends Theor. Comput. Sci.  \textbf{9}(3-4),  211--407 (2014)

\bibitem{hardt2016equality}
Hardt, M., Price, E., Srebro, N.: Equality of opportunity in supervised
  learning. Advances in neural information processing systems  \textbf{29},
  3315--3323 (2016)

\bibitem{jerant2000early}
Jerant, A.F., Johnson, J.T., Sheridan, C.D., Caffrey, T.J.: Early detection and
  treatment of skin cancer. American family physician  \textbf{62}(2),
  357--368 (2000)

\bibitem{kaissis2020secure}
Kaissis, G.A., Makowski, M.R., R{\"u}ckert, D., Braren, R.F.: Secure,
  privacy-preserving and federated machine learning in medical imaging. Nature
  Machine Intelligence  \textbf{2}(6),  305--311 (2020)

\bibitem{Kinyanjui20MICCAI}
Kinyanjui, N.M., Odonga, T., Cintas, C., Codella, N.C.F., Panda, R., Sattigeri,
  P., Varshney, K.R.: Fairness of classifiers across skin tones in dermatology.
  In: Martel, A.L., Abolmaesumi, P., Stoyanov, D., Mateus, D., Zuluaga, M.A.,
  Zhou, S.K., Racoceanu, D., Joskowicz, L. (eds.) Medical Image Computing and
  Computer Assisted Intervention -- MICCAI 2020. pp. 320--329. Springer
  International Publishing, Cham (2020)

\bibitem{larrazabal2020gender}
Larrazabal, A.J., Nieto, N., Peterson, V., Milone, D.H., Ferrante, E.: Gender
  imbalance in medical imaging datasets produces biased classifiers for
  computer-aided diagnosis. Proceedings of the National Academy of Sciences
  \textbf{117}(23),  12592--12594 (2020)

\bibitem{li2019fair}
Li, T., Sanjabi, M., Beirami, A., Smith, V.: Fair resource allocation in
  federated learning. arXiv preprint arXiv:1905.10497  (2019)

\bibitem{li2021estimating}
Li, X., Cui, Z., Wu, Y., Gu, L., Harada, T.: Estimating and improving fairness
  with adversarial learning. arXiv preprint arXiv:2103.04243  (2021)

\bibitem{li2020multi}
Li, X., Gu, Y., Dvornek, N., Staib, L.H., Ventola, P., Duncan, J.S.: Multi-site
  fmri analysis using privacy-preserving federated learning and domain
  adaptation: Abide results. Medical Image Analysis  \textbf{65},  101765
  (2020)

\bibitem{mmr+17}
McMahan, B., Moore, E., Ramage, D., Hampson, S., y~Arcas, B.A.:
  Communication-efficient learning of deep networks from decentralized data.
  In: Proceedings of Artificial Intelligence and Statistics (AISTATS). pp.
  1273--1282. PMLR (2017)

\bibitem{rieke2020future}
Rieke, N., Hancox, J., Li, W., Milletari, F., Roth, H.R., Albarqouni, S.,
  Bakas, S., Galtier, M.N., Landman, B.A., Maier-Hein, K., et~al.: The future
  of digital health with federated learning. NPJ digital medicine
  \textbf{3}(1), ~1--7 (2020)

\bibitem{saito2016s}
Saito, K., Yamada, H.: What’s so different about blockchain?—blockchain is
  a probabilistic state machine. In: 2016 IEEE 36th International Conference on
  Distributed Computing Systems Workshops (ICDCSW). pp. 168--175. IEEE (2016)

\bibitem{seyyedkalantari2020chexclusion}
Seyyed-Kalantari, L., Liu, G., McDermott, M., Chen, I.Y., Ghassemi, M.:
  Chexclusion: Fairness gaps in deep chest x-ray classifiers (2020)

\bibitem{simonyan2014very}
Simonyan, K., Zisserman, A.: Very deep convolutional networks for large-scale
  image recognition. arXiv preprint arXiv:1409.1556  (2014)

\bibitem{tschandl2018ham10000}
Tschandl, P., Rosendahl, C., Kittler, H.: The ham10000 dataset, a large
  collection of multi-source dermatoscopic images of common pigmented skin
  lesions. Scientific data  \textbf{5},  180161 (2018)

\bibitem{warnat2021swarm}
Warnat-Herresthal, S., Schultze, H., Shastry, K.L., Manamohan, S., Mukherjee,
  S., Garg, V., Sarveswara, R., H{\"a}ndler, K., Pickkers, P., Aziz, N.A.,
  et~al.: Swarm learning for decentralized and confidential clinical machine
  learning. Nature  \textbf{594}(7862),  265--270 (2021)

\bibitem{wu2019privacy}
Wu, Y., Yang, F., Xu, Y., Ling, H.: Privacy-protective-gan for privacy
  preserving face de-identification. Journal of Computer Science and Technology
   (2019)

\bibitem{ylct19}
Yang, Q., Liu, Y., Chen, T., Tong, Y.: Federated machine learning: Concept and
  applications. ACM Transactions on Intelligent Systems and Technology (TIST)
  \textbf{10}(2), ~12 (2019)

\bibitem{zhao2019secure}
Zhao, C., Zhao, S., Zhao, M., Chen, Z., Gao, C.Z., Li, H., Tan, Y.a.: Secure
  multi-party computation: theory, practice and applications. Information
  Sciences  \textbf{476},  357--372 (2019)

\end{thebibliography}
